\documentclass[prb,twocolumn,showpacs,preprintnumbers,amsmath,amssymb]{revtex4}
\usepackage{amscd,graphicx,color,colordvi,bm}

\newcommand{\vect}[1]{\mathbf{#1}}
\newcommand{\hvect}[1]{\hat{\mathbf{#1}}}

\renewcommand{\tensor}[1]{\bm{#1}}
\newcommand{\pdif}[2]{\frac{\partial #1}{\partial #2}}

\newcommand{\Fourier}[1]{\widetilde{#1}}

\newcommand{\betaP}{\beta^{\text{P}}}
\newcommand{\betaPtensor}{\tensor{\beta}^\text{P}}

\begin{document}

\preprint{cond-mat/0507456}

\title{%
  Mesoscale theory of grains and cells: crystal plasticity and coarsening
}

\author{Surachate Limkumnerd}
\email{sl129@cornell.edu}

\author{James P. Sethna}
\homepage{http://www.lassp.cornell.edu/sethna/sethna.html}
\affiliation{%
Laboratory of Atomic and Solid State Physics, Clark Hall,\\
Cornell University, Ithaca, NY 14853-2501, USA
}

\date{\today}

\begin{abstract}
  Solids with spatial variations in the crystalline axes naturally
  evolve into cells or grains separated by sharp walls. Such
  variations are mathematically described using the Nye dislocation
  density tensor. At high temperatures, polycrystalline grains form
  from the melt and coarsen with time: the dislocations can both climb
  and glide. At low temperatures under shear the dislocations (which
  allow only glide) form into cell
  structures. While both the 
  microscopic laws of dislocation motion and the macroscopic laws of
  coarsening and plastic deformation are well studied, we hitherto
  have had no simple, continuum explanation for the evolution of
  dislocations into sharp walls. We present here a
  mesoscale theory
  of dislocation motion. It provides a quantitative description of
  deformation and rotation, grounded in a microscopic order parameter
  field exhibiting the topologically conserved quantities. The
  topological current of the Nye dislocation density tensor is derived
  from a microscopic theory of glide driven by Peach-Koehler forces
  between dislocations using a simple closure approximation. The
  resulting theory is shown to form sharp dislocation walls in finite
  time, both with and without dislocation climb.
\end{abstract}


\maketitle

Crystals, when formed or deformed, relax by developing walls.  Common
metals (coins, silverware) are polycrystalline: the atoms locally
arrange into grains each with a specific crystalline lattice
orientation, separated by sharp, flat walls called \emph{grain
boundaries}. When metals are deformed (pounded or permanently bent) new
\emph{cell walls} form inside each
grain~\cite{Wilsdorf:1991,Hughes:1993,Hughes:1995}. Until now, our
only convincing 
understanding of why crystals form walls has been detailed and
microscopic. Our new theory provides an elegant, continuum description
of cell wall formation as the development of a \emph{shock front} - a
phenomenon hitherto associated with traffic jams and sonic booms.

Continuum theories of dislocation dynamics are not 
new~\cite{Eshelby:1956,Kroner:1958,Kosevich:1962,Mura:1963}. Various
approaches have been used to study dislocation pattern formation.
Simplified diffusion models have been used to describe
persistent slip bands~\cite{Walgraef:1985}, dislocation cell
structures during multiple slip~\cite{Hahner:1996}, and dislocation vein
structures~\cite{Saxlova:1997}. Models similar in spirit to ours have
recently been analyzed both for
single-slip~\cite{Groma:1997,Groma:1999,Zaiser:2001} and
in three dimensions~\cite{ElAzab:2000}. None of these approaches have
yielded the sharp, wall-like singularities characteristic of grain
boundaries and deformation-induced cell structures (see, however~\cite{Dawson:2002}).

Continuum plasticity theory, as used in practical engineering codes,
describes the deformation gradient 
($\partial_i u_j$, for displacement field $\vect{u}$) as the 
sum $\partial_i u_j = \beta_{ij}^\text{E} + \beta_{ij}^\text{P}$ of an
elastic, reversible distortion $\tensor{\beta}^\text{E}$ and a
permanent plastic 
distortion $\tensor{\beta}^\text{P}$. 
The plastic distortion involves the creation of microscopic
dislocations, and cannot be written as a gradient of a single-valued
displacement field. Integrating around a loop $L$
enclosing a surface $S$, the change in such a hypothetical plastic
distortion field $\Delta\vect{u}^\text{P}$ can be written using Stokes
theorem as 
\begin{equation}
  \Delta u^\text{P}_j = -\oint_L \betaP_{ij}\,dx_i = -\int_S
  \varepsilon_{ilm} \pdif{\betaP_{mj}}{x_l}\,dS_i = \int_S \rho_{ij}\,dS_i
\end{equation}
where the Nye dislocation density tensor
\begin{equation}\label{E:rho}
  \rho_{ij}(\vect{x}) = -\varepsilon_{ilm}\pdif{\betaP_{mj}}{x_l} =
  \sum_\alpha t^\alpha_i b^\alpha_j \delta(\vect{\xi}^\alpha)
\end{equation}
measures the net flux of dislocations $\alpha$, tangent to $\vect{t}$,
with Burgers vector $\vect{b}$, in the (coarse-grained) neighborhood
of $\vect{x}$. The microscopic statement that
dislocations cannot end implies
$\partial_i\rho_{ij} = 0$, so the time evolution must be
given in terms of a current~\cite{Kosevich:1962, Mura:1963,
  Rickman:1997} $\tensor{J}$: $\partial\rho_{ij}/\partial t =
-\varepsilon_{ilm}\partial J_{mj}/\partial x_l$.
From equation (\ref{E:rho})  we see that~\cite{vonMises}
\begin{equation}\label{E:EvolutionLaw}
  J_{ij} = \partial\betaP_{ij}/\partial t.
\end{equation} 

Thus the natural physicist's order parameter (the topologically
conserved dislocation density $\tensor{\rho}$) is a curl of the common
engineering state variable (the plastic distortion field
$\tensor{\beta}^\text{P}$). The focus of our manuscript will be the
derivation of an evolution law (\ref{E:EvolutionLaw}) appropriate for
scales large compared to the 
atoms but small compared to the cell structures and grain
boundaries.

Notice that we only consider the \emph{net} density of
dislocations. We ignore the
geometrically unneccessary dislocations (those with opposing Burgers
vectors which cancel out in the net dislocation 
density) because they do not affect the long range strain fields or
the misorientations at grain boundaries and cell
walls. Macroscopically they are known to dominate dislocation
entanglement and work hardening, and are included in previous
continuum theories of
plasticity~\cite{Zaiser:2001,Walgraef:1985,Pontes:2005,Groma:1997,Groma:1999,Selitser:1994}.
Much of the macroscopic cancellation 
in net dislocation density comes from the near alternation of the net
rotations in the series of cell walls~\cite{Hughes:PC}. Our focus on
the sub-cellular, subgrain length scales and our current omission of
dislocation tangling make keeping only the net dislocation density
natural for our purposes. We also do not explicitly incorporate a yield
surface, because we hope eventually to explain work hardening and
yield surfaces as properties which emerge from the intermediate
length-scale theory.

We motivate our 
evolution law from the microscopic Peach-Koehler force on a section of
a dislocation line due to the stress field $\tensor{\sigma}$  present
at that point: $f^\text{PK}_i = -\varepsilon_{ijk}t_j b_l \sigma_{kl}$.
The current $\tensor{J}$ of a single dislocation moving with velocity
$\vect{v}$ is $J_{ij} = \varepsilon_{ilm} t_l b_j v_m \delta(\vect{\xi})$.
If we treat dislocation glide and climb on an equal footing (a crude
model for metals at high temperatures), the dislocation will move in
the direction of the applied force $\vect{v}\propto
\vect{f}^\text{PK}$. For a single dislocation, tuning the
climb component with a parameter $\lambda$, we find
\begin{eqnarray}
  J_{ij} = &D\Big[&\varepsilon_{ilm} t_l b_j \varepsilon_{mpq}
  \sigma_{pr} t_q b_r \delta(\vect{\xi}) \nonumber \\
  &&-\frac{\lambda}{3}\,\delta_{ij} \varepsilon_{klm} t_l b_k
  \varepsilon_{mpq} \sigma_{pr} t_q b_r \delta(\vect{\xi})\Big].
\end{eqnarray}
Here $D$ is a materials constant with units of
$[\text{length}]^2[\text{time}]/[\text{mass}]$
giving the mobility of dislocation glide. At $\lambda = 0$ climb and
glide have equal mobilities, and at $\lambda = 1$ $\tensor{J}$ is
traceless and hence 
only glide is allowed. Coarse graining over many dislocations,
\begin{equation}
  J_{ij} = D_{ijkmpqrs}\sigma_{pq}\rho^{(4)}_{kmrs}
\end{equation}
where $D_{ijkmpqrs} =
D/2\big[\delta_{iq}\delta_{jm}\delta_{kr}\delta_{ps} -
  \delta_{ir}\delta_{jm}\delta_{kq}\delta_{ps} - \lambda/3\left(
    \delta_{ij}\delta_{mq}\delta_{kr}\delta_{ps} -
    \delta_{ij}\delta_{mr}\delta_{kq}\delta_{ps}\right)\big]$ is an
eighth rank isotropic tensor specific to 
Peach-Koehler model, and $\rho^{(4)}_{kmrs}$ is a higher order Nye
dislocation tensor. Thus that the evolution of $\tensor{\rho}$
depends upon the new object $\tensor{\rho}^{(4)}$.

We now need to perform a \emph{closure approximation}, writing 
$\tensor{\rho}^{(4)}$ in terms of $\tensor{\rho}$ (as in Hartree-Fock and in
theories of turbulence). The only choice $\tensor{\rho}^{(4)} \rightarrow
\tensor{\rho}\otimes\tensor{\rho}$ that guarantees a
decrease of elastic energy with time~\cite{Limkumnerd:TBP} is
\begin{eqnarray}
  \rho^{(4)}_{ijkm} &=& \sum_\alpha t^\alpha_i b^\alpha_j t^\alpha_k
  b^\alpha_m \delta(\vect{\xi}^\alpha) \\
  &\simeq& C\Big[\sum_\alpha
  t^\alpha_i b^\alpha_j \delta(\vect{\xi}^\alpha) \Big]\Big[\sum_{\alpha'}
  t^{\alpha'}_k b^{\alpha'}_m \delta(\vect{\xi}^{\alpha'}) \Big] =
  C\rho_{ij}\rho_{km},\nonumber
\end{eqnarray}
where $C$ has units of distance. The resulting evolution law for the plastic
distortion tensor is
\begin{eqnarray}\label{E:evolutionEq3D}
  \pdif{\betaP_{ij}}{t} = \frac{CD}{2}&\Big[&\left(\sigma_{ic}\rho_{ac}
      - \sigma_{ac}\rho_{ic}\right)\rho_{aj} \nonumber \\
    &&- \frac{\lambda}{3}\delta_{ij}\left(\sigma_{kc}\rho_{ac} -
  \sigma_{ac}\rho_{kc}\right)\rho_{ak}\Big]
\end{eqnarray}
where $\tensor{\rho}$ is a curl of $\betaPtensor$
(equation~\ref{E:rho}), and the stress $\tensor{\sigma}$ (due to the
long-range fields of the other dislocations) is written in Fourier
space as~\cite{Mura:1991}
\begin{eqnarray}
    \Fourier{\sigma}_{ij}(\vect{k}) =&
    K_{ijkl}(\vect{k})\Fourier{\rho}_{kl}(\vect{k}),&\\
    K_{ijkl}(\vect{k}) =& -\dfrac{i\mu
      k_m}{k^2}\Big[\varepsilon_{ilm}\delta_{jk} &+
    \varepsilon_{jlm}\delta_{ik} \nonumber \\
    &&+ \frac{2\varepsilon_{klm}}{1-\nu}\left(\frac{k_i k_j}{k^2} -
      \delta_{ij}\right) \Big].\nonumber
\end{eqnarray}

For simplicity, we present solutions to our three-dimensional theory
for the special case of systems where the dislocation density varies
only along one and two dimensions. In one dimension with variations
only along $\hvect{z}$, our theory becomes local, with stress
\begin{equation}
  \tensor{\sigma}(z) = -\mu\begin{pmatrix}
    \dfrac{2}{1-\nu}(\betaP_{xx} + \nu\betaP_{yy}) & \betaP_{xy} +
    \betaP_{yx} & 0 \\
    \betaP_{xy} + \betaP_{yx} & \dfrac{2}{1-\nu}(\nu\betaP_{xx} +
    \betaP_{yy}) & 0 \\
    0 & 0 & 0
  \end{pmatrix},
\end{equation}
Nye dislocation tensor
\begin{equation}
  \rho_{xj} = \partial_z \betaP_{yj},\qquad \rho_{yj} =
  -\partial_z\betaP_{xj}, \qquad \rho_{zj} = 0,
\end{equation}
and evolution law
\begin{equation}\label{E:evolutionEq}
  \pdif{\betaPtensor}{t} =
  -F(\betaPtensor)\pdif{\betaPtensor}{z}
\end{equation}
with
\begin{eqnarray}
  F(\betaPtensor) &=
  -\pdif{w(\betaPtensor)}{z} 
  = -CD\mu \pdif{}{z}\Big[\frac{1}{2}\left(\betaP_{xy} +
    \betaP_{yx}\right)^2 \nonumber\\
  &+ \left({\betaP_{xx}}^2 + {\betaP_{yy}}^2 \right) +
  \frac{\nu}{1-\nu}\left(\betaP_{xx}+\betaP_{yy} \right)^2
  \Big].
\end{eqnarray}
Interestingly, $w(\betaPtensor)$ is the linear elastic
energy density of the 
system, the effective force $F(\betaPtensor)$ is
proportional to the $z$-component
of the Peach-Koehler force, and the rate of change of the elastic
energy density is simply  $-F^2(\betaPtensor)$ (so the
energy decreasing condition is explicit).

Our equation for $\betaPtensor$ in one dimension is
hyperbolic: the method of characteristics may be applied, where the
characteristics are the parameterized 
curves $(t(s), z(s))$ with $dt/ds = 1, dz/ds = F(\betaPtensor)$.
Along these curves,
\begin{equation}
  \frac{d\betaPtensor}{ds} = \frac{dt}{ds}\pdif{\betaPtensor}{t} +
  \frac{dz}{ds}\pdif{\betaPtensor}{z} = \pdif{\betaPtensor}{t} +
  F(\betaPtensor)\pdif{\betaPtensor}{z} = 0,
\end{equation}
implying (from equation~\ref{E:evolutionEq}) that $\betaPtensor$ is
constant along the characteristics. As with many hyperbolic systems,
we observe shock formuation in finite time. Our analytic analysis of
the shock asymptotics (in preparation) exhibits the jump in
$\betaPtensor$, wall in $\tensor{\rho}$, and stress jump
described numerically below, with a $1/\sqrt{t}$ time-dependence for
the surrounding continuum $\betaPtensor$.

  We simulate systems of spatial extent $L$ in one or two dimensions,
  with periodic boundary conditions and no external stress. The
  initial plastic distortion field $\betaPtensor$ is a Gaussian random
  field with mean square amplitude $\betaP_0$ and root-mean-square decay length
  approximately $L/16$. In one dimension, we use the upwind
  scheme~\cite{Shu:1989} as 
  implemented by Press \emph{et al}~\cite{Press:2002}. In two dimensions, we
  calculate $\tensor{\sigma}$ using 
  Fourier methods and regularize our equations by adding a
  fourth-order numerical viscosity $-K\nabla^4\betaP_{ij}$ to
  equation~\ref{E:evolutionEq3D}. (A second-order
  viscosity $\nabla^2 \betaP_{ij}$ was not as successful in
  suppressing the instabilities.) 
  In one dimension, we have checked that the upwind scheme produces
  the same solution as the viscosity regularization. Unlike shock
  waves in fluids, where conservation laws tell us that the 
  upwind/viscosity solution is the correct one, here the dynamics of
  the walls is not determined by the continuum theory. (We speculate
  that the wall dynamics regularization may be non-trivial and
  stochastic, emerging out of avalanches and critical depinning
  phenomena on sub-cell length scales -- observed both
  experimentally~\cite{Richeton:2005} and
  theoretically~\cite{Miguel:2002} in systems with only one 
  active slip system.)

  \begin{figure}[htb]
    \centering
    \includegraphics[width=.5\textwidth]{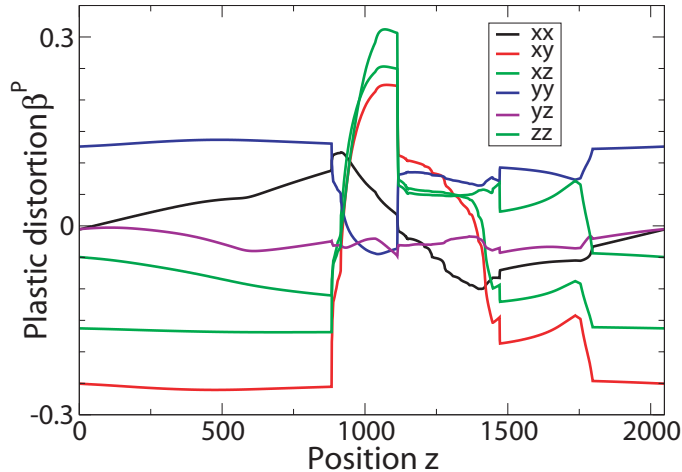}
    \caption{Plastic distortion tensor component $\betaP_{ij}$ in one
    dimension allowing only glide motion, after time $t=20L^2/D\mu$,
    with 2048 mesh points. The shocks or jumps in the values
    correspond to the cell walls. A stress-free wall (satisfying
    the Frank conditions) here would have no jump in 
    $\betaP_{xx}$, $\betaP_{yy}$, or $\betaP_{xy}+\betaP_{yx}$.}
    \label{fig:betaPglide}
  \end{figure}
  Figure~\ref{fig:betaPglide} shows the final state of the plastic distortion
  $\betaPtensor$ in a 
  one-dimensional system evolving under dislocation glide only
  (corresponding to plastic deformation, which experimentally
  leads to cell structure formation). Jump singularities in $\betaPtensor$ form
  after a short time, representing a wall of dislocations and an
  abrupt change in lattice orientation. The cell walls in our model
  are not grain boundaries, because they do not correspond to pure,
  simple rotations (they do not satisfy the Frank
  conditions~\cite{Read:1953}). There
  is a jump in the stress across each cell wall: the glide component
  of the forces from the two neighboring cells on $\tensor{\rho}$ in a wall is
  equal, opposite, and compressive. Our model therefore predicts that
  the initial formation of cell walls during low-temperature plastic
  deformation is driven by a stress jump due to non-cancellation of
  the stress field of the constituent dislocations. As our system
  evolves, new cell 
  walls form (intriguingly similar to the cell refinement seen
  experimentally~\cite{Hughes:1998,Sethna:2003}) and the
  existing cell walls evolve to reduce their stress jumps.

  \begin{figure}[htb]
    \centering
    \includegraphics[width=.5\textwidth]{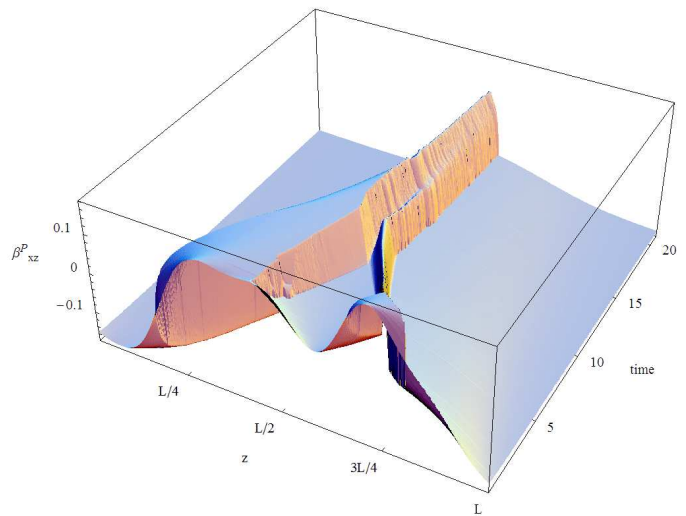}
    \caption{The $xz$-component of the plastic distortion tensor in
    one dimension up to time $t=22L^2/D\mu$, with 2048 mesh
    points. The evolution allows both glide and climb motions. The
    walls move and coalesce until only a single wall survives.}
    \label{fig:betaPglideclimb}
  \end{figure} 
  Figure~\ref{fig:betaPglideclimb} shows the evolution of the plastic
  distortion in a system 
  allowing both glide and climb, corresponding to high-temperature
  annealing of a plastically deformed structure. Here the initial
  singularities again are not grain boundaries, but walls of
  dislocations with net stress jumps. These walls then move and coalesce in
  response to the net force due to other dislocations,
  coarsening the resulting grain boundary structure.

  \begin{figure}[htb]
    \centering
    \includegraphics[width=.5\textwidth]{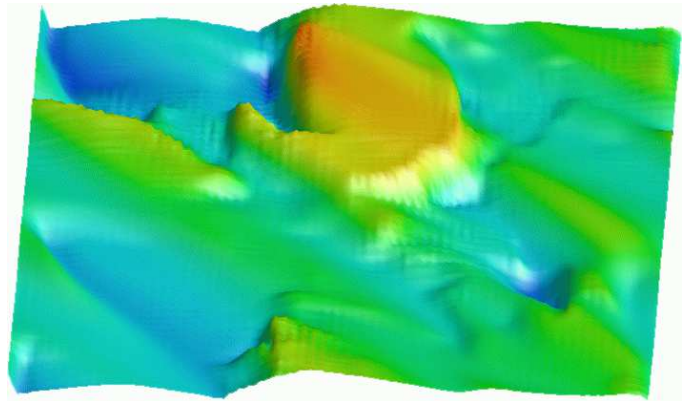}
    \caption{The $yz$-component of the plastic distortion tensor
    allowing only glide, in two dimensions after time $t=9.15 L^2/D\mu$.}
    \label{fig:beta2D}
  \end{figure}
  Figure~\ref{fig:beta2D} shows the final state of $\betaPtensor$ in a
  two-dimensional plasticity
  simulation (with glide and not climb). Notice the formation of sharp
  walls (mathematically demonstrated above only in one dimension).
  These walls separate relatively unstrained regions which we identify
  with the cells formed during plastic deformation. As in the
  one-dimensional simulations, the cell walls formed at short times
  have stress jumps.

  Generically we expect our model to form sharp walls in finite
  time. Special initial conditions, however, may not form walls. Indeed,
  the case of a single slip system does not form singularities within
  our model---perhaps explaining why cell wall formation is not observed 
  in Stage~I hardening, or in dislocation-dynamics studies of single
  slip in two dimensions otherwise similar to our approach~\cite{Groma:1997,Groma:1999,Zaiser:2001}.
  
Our dislocation dynamics theory is minimalist: it ignores many
  features (geometrically unnecessary dislocations~\cite{Zaiser:2001},
  slip systems~\cite{Mika:1999},
  dislocation tangling, yield surfaces, nucleation of new
  dislocations) that are known 
  to be macroscopically important in real materials. It does
  incorporate cleanly and 
  microscopically the topological constraints, long-range forces and
  energetics driving the dislocation dynamics. As hypothesized by the
  LEDS (low-energy dislocation structures)
  approach~\cite{Wilsdorf:1989,Wilsdorf:1995}, we find that a dynamics
  driven by minimizing energy (omitting
  tangling and nucleation) still produces cell boundary
  structures. Finally our theory, to our surprise, initially forms
  sharp walls that are not the usual zero-stress grain boundaries.

  In condensed-matter physics, crystals are anomalous. Most phases
  (liquid crystals, superfluids, superconductors, magnets) respond
  smoothly in space when strained. Crystals, when rotationally strained,
  form sharp walls separating cells or grains. Our analysis suggests
  that the formation of these walls can be understood as a shock
  formation in the dislocation dynamics. 

\begin{acknowledgements}
  We would like to thank Jean-Philippe Bouchaud, Valerie Coffman, Paul
  Dawson, Matthew Miller, Markus Rauscher, Wash Wawrzynek, and Steven
  Xu for helpful instruction on plasticity and numerical methods, and
  funding from NSF grants ITR/ASP ACI0085969 and DMR-0218475.
\end{acknowledgements}

\bibliography{references}

\end{document}